\documentclass[twocolumn,floats,rmp]{revtex4}
\usepackage{graphics,graphicx,epsfig}
\usepackage{amssymb}
\usepackage{epsf,epstopdf,wrapfig}
\usepackage[]{natbib}

\begin{document}

\title{Cell biology: Networks, regulation, pathways}

\author{Ga\v{s}per Tka\v{c}ik  and William Bialek$^*$}
\affiliation{Joseph Henry Laboratories of Physics,  Lewis--Sigler Institute for Integrative Genomics, and $^*$Princeton Center for Theoretical Physics, Princeton University, Princeton, New Jersey 08544}

\date{\today}

\begin{abstract}
This review was written for the {\em Encyclopedia of Complexity and System Science} (Springer--Verlag, Berlin, 2008), and is intended as a guide to the growing literature which approaches the phenomena of cell biology from a more theoretical point of view.  We begin with the building blocks of cellular networks, and proceed toward the different classes of models being explored, finally discussing the `design principles' which have been suggested for these systems.  Although largely a dispassionate review, we do draw attention to areas where there seems to be general consensus on ideas that have not been tested very thoroughly and, more optimistically, to areas where we feel promising ideas deserve to be more fully explored.
\end{abstract}

\maketitle 
\tableofcontents

\section{Introduction}
\emph{Biological network} has come to mean a
system of interacting molecules that jointly perform  cellular tasks such as the regulation of gene expression, information transmission, or metabolism \citep{Bray1995a}. Specific instances of biological networks include, for example, the DNA and DNA binding proteins comprising the transcriptional regulatory network; signaling proteins and small molecules comprising various signaling networks; or enzymes and metabolites comprising the metabolic network. Two important assumptions shape our current understanding of such systems: first, that the biological networks have been under selective evolutionary pressure to perform specific cellular functions in a way that furthers the overall reproductive success of the individual; and second, that these functions often are not implemented on a microscopic level by single molecules, but are rather a collective property of the whole interaction network. The question of how complex behavior emerges in a network of (simple) nodes under a functional constraint is thus central \citep{Strogatz2001a}.

To start off with a concrete example, consider chemotaxis in the bacterium  \emph{Escherichia coli}  \citep{Berg1975a, Falke1997a}, one of the paradigmatic examples of signal transduction. This system is dedicated to steering the bacteria towards areas high in nutrient substances and away from repellants.   Chemoeffector molecules in the solution outside the bacterium bind to receptor molecules on the cell surface, and the resulting structural changes in the receptors are relayed in turn by the activities of the intracellular signaling proteins to generate a control signal for molecular motors that drive the bacterial flagella.  The chemotactic network consists of about 10 nodes (here, signaling proteins), and the interactions between the nodes are the chemical reactions of methylation or phosphorylation. 
Notable features of this system include its extreme sensitivity, down to the limits set by counting individual molecules as they arrive at the cell surface \citep{Berg1977a}, and the maintenance of this sensitivity across a huge dynamic range, through an adaptation mechanism that provides nearly perfect compensation of  background concentrations \citep{Block1983a}.  More recently it has been appreciated that aspects of this functionality, such as perfect adaptation, are also robust against large variations in the concentrations of the network components \citep{Barkai1997a}.

Abstractly, different kinds of signaling proteins, such those in chemotaxis, can be thought of as the building blocks of a network, with their biochemical interactions forming the wiring diagram of the system, much like the components and wiring diagram of, for instance, a radio receiver. 
In principle, these wiring diagrams are hugely complex; for a network composed of $N$ species, there are $\sim C^N_k$ possible connections among any set of $k$ components, and typically we don't have direct experimental guidance about the numbers associated with each `wire.'   One approach is to view this a giant fitting problem:  once we draw a network, there is a direct translation of this graph into dynamical equations, with many parameters, and we should test the predictions of these dynamics against whatever data are available to best determine the underlying parameters.  Another approach is to ask whether this large collection of parameters is special in any way other than that it happens to fit the data--are there principles that allow us to predict how these systems \emph{should} work?  In the context of chemotaxis, we might imagine that network parameters have been selected to optimize the average progress of bacteria up the chemical gradients of nutrients, or to maximize the robustness of certain functions against extreme parameter variations.  These ideas of design principles clearly are not limited to bacterial chemotaxis.

An important aspect of biological networks is that the same  components (or components that have an easily identifiable evolutionary relationship) can be (re)used in different modules or used for the same function in a different way across species, as discussed for example by  \citet{Rao2004a} for the case of bacterial chemotaxis. Furthermore, because evolutionary selection depends on function and not directly on microscopic details,  different wiring diagrams or even changes in components themselves can result in the same performance; evolutionary process can gradually change the structure of the network as long as its function is preserved; as an example see the discussion of transcriptional regulation in yeast by  \citet{Tanay2005a}.
On the other hand, one can also expect that signal processing problems like gain control, noise reduction, ensuring (bi)stability etc, have appeared and were solved repeatedly, perhaps even in similar ways across various cellular functions, and we might be able to detect the traces of their commonality in the network structure, as for example in the discussion of local connectivity in bacterial transcriptional regulation by  \citet{Shen-Orr2002a}. Thus there are reasons to believe that in addition to design principles at the network level, there might also be local organizing principles, similar to common wiring motifs in electronic circuitry, yet still independent of the identity of the molecules that implement these principles.

Biological networks have been approached at many different levels, often by investigators from different disciplines.  The basic wiring diagram of a network---the fact that a kinase phosphorylates these particular proteins, and not all others, or that a transcription factor binds to the promoter regions of particular genes---is determined by classical biochemical and structural concepts such as binding specificity.  At the opposite extreme, trying to understand the collective behavior of the network as a whole suggests approaches from statistical physics, often looking at simplified models that leave out many molecular details.  Analyses that start with design principles are yet a different approach, more in the `top--down' spirit of statistical physics but leaving perhaps more room for details to emerge as the analysis is refined.  Eventually, all of these different views need to converge:  networks really are built out of molecules, their functions emerge as  collective behaviors, and these functions must really be functions of use to the organism.  At the moment, however, we seldom know enough to bridge the different levels of description, so the different approaches are pursued more or less independently, and we follow this convention here.  We will start with the molecular building blocks, then look at models for networks as a whole, and finally consider design principles.  We hope that this  sequence doesn't leave the impression that we actually know how to build up from molecules to function!

Before exploring our subject in more detail, we take a moment to consider its boundaries.  Our assignment from the editors was to focus on phenomena at the level of molecular and cellular biology.  A very different approach attempts to create a `science of networks' that searches for common properties in biological, social, economic and computer networks \citep{Newman2006}.  Even within the biological world, there is a significant divide between work on networks in cell biology and networks in the brain.  As far as we can see this division is an artifact of history, since there are many issues which cut across these different fields.  Thus, some of the most beautiful work on signaling comes from photoreceptors, where the combination of optical inputs and electrical outputs allowed, already in the 1970s, for experiments with a degree of quantitative analysis that even today is hard to match in systems which take chemical inputs and give outputs that modulate the expression levels of genes \cite{Baylor1979b, Rieke1998a}.  Similarly, problems of noise in the control of gene expression have parallels in the long history of work on noise in ion channels, as we have discussed elsewhere \citep{Tkacik2007c}, and the problems of robustness have also been extensively explored in the network of interactions among the multiple species of ion channels in the membrane \cite{LeMasson1993a, Goldman2001a}.  Finally, the ideas of collective behavior are much better developed in the context of neural networks than in cellular networks, and it is an open question how much can be learned by studying these different systems in the same language \cite{Tkacik2007d}.

\section{Biological networks and their building blocks}
\subsection{Genetic regulatory networks}
Cells constantly adjust their levels of gene expression.
One central mechanism in this regulatory process involves the control of transcription by proteins known as
transcription factors (TFs), which locate and bind short DNA sequences in the regulated genes' promoter or enhancer  regions. A given transcription factor can regulate either a few or a sizable proportion of the genes in a genome, and a single gene may be regulated by more than one transcription factor; different transcription factors can also regulate each other \citep{Watson2003a}.

In the simplest case of a gene regulated by a single TF, the gene might be expressed whenever the factor -- in this case called an  activator -- is bound to the cognate sequence in the promoter (which corresponds to the situation when the TF concentration in the nucleus is high), whereas the binding of a repressor would shut a normally active gene down. 
The outlines of these basic control principles were established long ago, well before the individual transcription factors could be isolated, in elegant experiments on  the \emph{lactose} operon of \emph{Escherichia coli} 
\citep{Jacob1961a} and even simpler model systems such as phage $\lambda$ \citep{Ptashne2004a}.  To a great extent the lessons learned from these experiments have provided the framework for understanding transcriptional control more generally, in prokaryotes \citep{Ptashne2001a}, eukaryotes \citep{Kadonaga2004a}, and even during the development of complex multicellular organisms \citep{Arnosti2005a}.
 
The advent of high throughput techniques for probing gene regulation has extended our reach beyond single genes. In particular, microarrays \citep{Brown1999a} and the related data analysis tools, such as clustering \citep{Eisen1998a}, have enabled researchers to find sets of genes, or \emph{modules}, that are \emph{coexpressed}, i.e. up- or down-regulated in a correlated fashion when the organism is exposed to different external conditions, and are thus probably regulated by the same set of transcription factors. Chromatin immunoprecipitation (ChIP) assays have made it possible to directly screen for short segments of DNA that known TFs bind; using microarray technology it is then possible to locate the intergenic regions which these segments belong to, and hence find the regulated genes, as has recently been done for the \emph{Saccharomyces cerevisiae} DNA-TF interaction map \citep{Lee2002a}.
 
These high throughput experimental approaches, combined with traditional molecular biology and complemented by sequence analysis and related mathematical tools  \cite{Siggia2005a}, provide a large scale, topological view of the transcriptional regulatory network of a particular organism, where each link between two nodes (genes) in the  regulatory graph implies either activation or repression \cite{Alm2003a}. While useful for describing causal interactions and trying to predict responses to mutations and external perturbations \citep{Levine2005a}, this picture does not explain how the network operates on a physical level: it lacks dynamics and specifies neither the strengths of the interactions nor how all the links converging onto a given node jointly exercise control over it. To address these issues, representative wild-type or simple synthetic regulatory elements and networks consisting of a few nodes have been studied extensively to construct quantitative models of the network building blocks.

For instance, combinatorial regulation of a gene by several transcription factors that bind and interact on the promoter has been considered by \citet{Buchler2003a} as an example of (binary) biological computation and synthetic networks implementing such computations have been created \cite{Guet2002a, Yokobayashi2002a}. Building on classical  work describing allosteric proteins such as hemoglobin,  thermodynamic models have been used with success to account for combinatorial interactions on the operator of the lambda phage \cite{Ackers1982a}.  More recently \citet{Bintu2005a, Bintu2005b} have reviewed the equilibrium statistical mechanics of such interactions,  \citet{Setty2003a} have experimentally and systematically mapped out the response surface of the \emph{lac} promoter to combinations of its two regulatory inputs, cAMP and IPTG, and \citet{Kuhlman2007a} have finally provided a consistent picture of the known experimental results and the thermodynamic model for the combinatorial regulation of the lactose operon. There have also been some successes in eukaryotic regulation, where \citet{Schroeder2004a} used thermodynamically motivated models to detect  clusters of binding sites that regulate the gap genes in morphogenesis of the fruit fly.

Gene regulation is  a dynamical process composed of a number of steps, for example the binding of TF to DNA, recruitment of transcription machinery and the production of the messenger RNA, post-transcriptional regulation, splicing and transport of mRNA, translation, maturation and possible localization of proteins. While the extensive palette of such microscopic interactions represents a formidable theoretical and experimental challenge for each detailed study, on a network level it primarily induces three effects. First, each node -- usually understood as the amount of gene product -- in a graph of regulatory interactions is really not a single dynamical variable and thus  has some internal state, representing the configuration on the associated promoter, concentration of the corresponding messenger RNA etc.; the relation of these quantities to the concentration of the output protein is not necessarily straightforward, as emphasized in recent work comparing   mRNA and protein levels in yeast \citep{Ghaemmaghami2003a}.  Second, collapsing multiple chemical species onto a single node makes it difficult to include 
non--transcriptional regulation of gene expression in the same framework. Third, the response of the target gene to changes in the concentrations of its regulators will be delayed and extended in time, as in the example explored by \citet{Rosenfeld2003a}. 

Perhaps the clearest testimonies to the importance of dynamics in addition to network topology are provided by systems that involve regulatory loops, in which the output of a network feeds back on one of the inputs as an activator or repressor. \citet{McAdams1995a} have argued that the time delays in genetic regulatory elements are essential for the proper functioning of the  phage $\lambda$ switch, while \citet{Elowitz2000a} have created a synthetic circuit made up of three mutually repressing genes (the ``repressilator''), that exhibits spontaneous oscillations. Circadian clocks  are examples of naturally occurring genetic oscillators \citep{Young2001a}.

In short, much is known about the skeleton of genetic regulatory interactions for model organisms, and physical models exist for several well studied (mostly prokaryotic) regulatory elements. While homology allows us to bridge the gap between model organisms and their relatives, it is  less clear how and at which level of detail  the knowledge about regulatory elements must be combined into a network to explain and predict its function.
\subsection{Protein--protein interaction networks}
After having been produced, proteins often assemble into complexes through direct contact  interactions, and these complexes are functionally active units participating in signal propagation and other pathways. Proteins also interact through less persistent encounters, as when a protein kinase meets its substrate.  It is tempting to define a link in the network of protein--protein interactions by such physical associations, and this is the basis of several experimental methods which aim at a genome--wide survey of of these interactions.
Although starting out being relatively unreliable (with false positive rates of up to $50\%$), high throughput techniques like the yeast two hybrid assay \citep{Uetz2000a,Ito2001a} or mass spectrometry \citep{Gavin2002a,Ho2002a} are providing data of increasing quality about protein--protein interactions, or the ``interactome'' \citep{Krogan2006a}.  While more reliable methods are being developed  \citep{Alm2003a} and new organisms are being analyzed in this way \citep{Giot2003a, Li2004a, Rual2005a}, the existing interaction data from high throughput experiments and curated databases has already been extensively studied.

Interpretation of the interactions in the protein network is tricky, however, due to the fact that different experimental approaches have various biases -- for example, mass spectrometry is biased towards detecting interactions between proteins of high abundance, while two hybrid methods seem to be unbiased in this regard; on the other hand, all methods show some degree of bias towards different cellular localizations and evolutionary novelty of the proteins. Assessing such biases, however, currently depends not on direct calibration of the methods themselves but on comparison of the results with manually curated databases; although the databases surely have their own biases  \citep{Jansen2004a}. It is reassuring that the intersection of various experimental results shows significantly improved agreement with the databases, but this comes at the cost of a substantial drop in coverage of the proteome \citep{vonMering2002a}.

In contrast to the case of transcriptional regulation, the relationship between two interacting proteins is symmetric: if protein A binds to protein B, B also binds to A, so that the network is described by an undirected graph.
Most of the studies have been focused on binary interactions that yeast two hybrid and derived approaches can probe, although spectrometry can detect multiprotein complexes as well. Estimates of number of links in these networks vary widely, even in  the yeast \emph{Saccharomyces cerevisiae}: \citet{Krogan2006a} directly measure around 7100 interactions (between 2700 proteins), while \citet{Tucker2001a} estimate the total to be around 13000--17000, and \citet{vonMering2002a} would put the lower estimate at about 30000. Apart from the experimental biases  that can influence such estimates and have been discussed already, it is important to realize that each experiment can only detect interactions between proteins that are expressed under the chosen external conditions (e.g. the nutrient medium); moreover, interactions can vary from being transient to permanent, to which various measurement methods respond differently. It will thus become increasingly important to qualify each interaction in a graph by specifying how it depends on context in which the interaction takes place.

Proteins ultimately carry out most of the cellular processes such as transcriptional regulation, signal propagation and metabolism, and these processes can be modeled by their respective network and dynamical system abstractions. In contrast, the interactome is not a dynamical system itself, but instead captures specific reactions (like  protein complex assembly) and structural and/or functional relations that are present in all of the above processes. In this respect it has an important practical role of annotating currently unknown proteins through `guilt by association', by tying them into complexes and processes with a previously known function.
\subsection{Metabolic networks}
Metabolic networks organize our knowledge about anabolic and catabolic reactions between the enzymes, their substrates and co-factors (such as ATP), by reducing the set of reactions to a graph representation where two substrates are joined by a link if they participate in the same reaction. For model organisms like the bacterium \emph{Escherichia coli} the metabolic networks have been studied in depth and are publicly available \citep{Kanehisa2002a, Karp2002a}, and an increasing number of analyzed genomes offers sufficient sampling power to make statistical statements about the network properties across different domains of life \citep{Jeong2000a}. 

Several important features distinguish metabolic from protein--protein interaction and transcriptional regulation networks. First, for well studied systems the coverage of metabolic reactions is high, at least for the central routes of energy metabolism and small molecule synthesis; notice that this is a property of our knowledge, not a property of the networks (!). Second, cellular concentrations of metabolites usually are much higher than those of transcription factors, making the stochasticity in reactions due to small molecular counts irrelevant. Third, knowledge of the stoichiometry of reactions allows one to directly write down a system of first order differential equations for the metabolite fluxes \citep{Heinrich1996a}, which in steady state reduces to a set of linear constraints on the space of solutions. These chemical constraints go beyond topology and can yield strong and testable predictions; for example, \citet{Ibarra2002a} have shown how computationally maximizing the growth rate of \emph{Escherichia coli} within the space of allowed solutions given by flux balance constraints can correctly predict measurable relationships between oxygen and substrate uptake, and that bacteria can be evolved towards the predicted optimality for growth conditions in which the response was initially suboptimal. 
\subsection{Signaling networks}
Signaling networks consist of receptor and signaling proteins that integrate, transmit and route information by means of chemical transformations of the network constituents. One class of such transformations, for example, are  post--translational modifications, where targets are phosphorylated, methylated, acetylated, $\dots$ on specific residues, with a resulting change in their enzymatic (and thus signaling) activity. Alternatively, proteins might form stable complexes or dissociate from them, again introducing states of differential activity. The ability of cells to modify or tag proteins (possibly on several residues) can increase considerably the cell's capacity to encode its state and transmit information, assuming that the signaling proteins are highly specific not only for the identity but also the modification state of their targets; for a review see \citet{Papin2005a} .

Despite the seeming overlap between the domains of protein--protein network and signaling networks, the focus of the analysis is substantially different.  The interactome is simply a set of  possible protein--protein interactions and thus a topological (or connectivity) map; in contrast, signaling networks aim to capture signal transduction and therefore need to establish a causal map, in which the nature of the protein--protein interaction, its direction and timescale, and its quantitative effect on the activity of the target protein matter.  As an example, see this discussion by  \citet{Kolch2000a} on the role of protein--protein interactions in MAPK signaling cascade. 

Experiments on some signaling systems, such as the \emph{Escherichia coli} chemotactic module, have generated enough experimental data to require detailed models in the form of dynamical equations. Molecular processes in a signaling cascade extend over different time scales, from milliseconds required for kinase and phosphatase reactions and protein conformational changes, to minutes or more required for gene expression control, cell movement and receptor trafficking; this fact, along with the (often essential) spatial effects such as the localization of signaling machinery and diffusion of chemical messengers, can considerably complicate analyses and simulations.

Signaling networks are often factored into pathways that have specific inputs, such as the ligands of the G protein coupled receptors on the cell surface, and specific outputs, as with pathways that   couple to the transcriptional regulation apparatus or to changes in the
intracellular concentration of messengers such as calcium or cyclic nucleotides.  Nodes in signaling networks can participate in several pathways simultaneously, thus enabling signal integration or potentially inducing damaging  ``crosstalk'' between pathways; how junctions and nodes process signals is an area of active research \cite{Jordan2000a}. 

The components of signaling networks have long been the focus of biochemical research, and genetic methods allow experiments to assess the impact of knocking out or over--expression particular components.
In addition, several experimental approaches are being designed specifically for elucidating signaling networks. Ab--chips localize various signaling proteins on chips reminiscent of DNA microarrays, and stain them with appropriate fluorescent antibodies \cite{Nielsen2003a}. Multicolor flow cytometry is performed on cells immuno-stained for signaling protein modifications and hundreds of single cell simultaneous measurements of the modification state of pathway nodes are collected \cite{Perez2002a}.  Indirect inference of signaling pathways is also possible from genomic or proteomic data. 

One well studied signal transduction system is the mitogen activated protein kinase (MAPK) cascade that controls, among other functions, cell proliferation and differentiation \cite{Chang2001a}. Because this system is present in all eukaryotes and its structural components are used in multiple pathways, it has been chosen as a paradigm for the study of specificity and crosstalk. Similarly, the TOR system, identified  initially in yeast, is responsible for integrating the information on nutrient availability, growth factors and energy status of the cell and correspondingly regulating the cell growth \cite{Martin2005a}. Another interesting example of signal integration and both intra- and inter-cellular communication is observed in the quorum sensing circuit of the bacterium \emph{Vibrio harveyi}, where different kinds of species- and genus-specific signaling molecules are detected by their cognate receptors on the cell surface, and the information is fed into a common \emph{Lux} phosphorelay pathway which ultimately regulates the quorum sensing genes \cite{Waters2005a}.
\section{Models of biological networks}
\subsection{Topological models}
The structural features of a network are captured by its connectivity graph, where interactions (reactions, structural relations) are depicted as the links between the interacting nodes (genes, proteins, metabolites). Information about connectivity clearly cannot and does not describe the network behavior, but it might influence and constrain it in revealing ways, similar to effect that the topology of the lattice has on the statistical mechanics of systems living on it.

Theorists have studied extensively the properties of regular networks and random graphs starting with Erd\"{o}s and R\'{e}nyi in 1960s. The first ones are characterized by high symmetry inherent in  a square, triangular, or all-to-all (mean field) lattice; the random graphs were without such regularity, constructed simply by distributing $K$ links at random between $N$ nodes. 
The simple one--point statistical characterization that distinguishes random from regular networks looks at the node degree, that is the probability $P(k)$ that any node has $k$ incoming and/or outgoing links. For random graphs this distribution is Poisson, meaning that most of the nodes have degrees very close to the mean, $\langle k \rangle = \sum_k k\,P(k)$, although there are fluctuations;  for regular lattices every node has the same connectivity  to its neighbors.

The first analyses of the early reconstructions of large metabolic networks revealed a surprising ``scale free'' node degree distribution, that is 
$P(k)\sim k^{-\gamma}$, with $\gamma$ between 2 and 3 for most networks.
For the physics community, which had seen the impact of such scale invariance on our understanding of phase transitions, these observations were extremely suggestive. It should be emphasized that for many problems in areas as diverse as quantum field theory, statistical mechanics and dynamical systems, such scaling relations are much more than curiosities.  Power laws relating various experimentally observable quantities are exact (at least in some limit), and the exponents (here, $\gamma$) really contain everything one might want to know about the nature of order in the system.  
Further, some of the first thoughts on scaling emerged from phenomenological analyses of real data.
Thus, the large body of work on scaling ideas in theoretical physics set the stage for people to be excited by the experimental observation of power laws in much more complex systems, although it is not clear to us whether the implied promise of connection to a deeper theoretical structure has been fulfilled. For divergent views on these matters see \citet{Barabasi2002} and \citet{Keller2005a}.

The most immediate practical consequence of a scale free degree distribution is  that---relative to expectations based on random graphs---there will be an over--representation of nodes with very large numbers of links, as with pyruvate or coenzyme A in metabolic networks \cite{Jeong2000a, Wagner2001a}.  These are sometimes called hubs, although another consequence of a scale free distribution is that there is no `critical degree of connection' that distinguishes hubs from non--hubs.
In the protein--protein interaction network of \emph{Saccharomyces cerevisiae}, nodes with higher degree are more likely to represent essential proteins \citep{Jeong2001a}, suggesting that node degree does have some biological meaning.  On the theoretical side,
removal of a sizeable fraction of nodes from a scale free network will neither increase the network diameter much, nor partition the network into equally sized parts \cite{Albert2000a}, and it is tempting to think that this robustness is also biologically significant.  The scale free property has been observed in many non-biological contexts, such as the topology of social interactions, World Wide Web links, electrical power grid connectivity ...  \cite{Strogatz2001a}.
A number of models have been proposed for how such scaling might arise, and some of these ideas, such as growth by preferential attachment, have a vaguely biological flavor \cite{Barabasi1999a, Barabasi2004a}.  Finding the properties of networks that actually discriminate among different mechanisms of evolution or growth turns out to be surprisingly subtle \cite{Ziv2005a}.

Two other revealing measures are regularly computed for biological networks. The mean path length,  $\langle l \rangle$, is the shortest path between a pair of nodes, averaged over all pairs in the graph, and  measures  the network's overall `navigability.'   Intuitively, short path lengths correspond to, for example, efficient or fast flow of information and energy in signaling or metabolic networks, quick spread of  diseases in a social network and so on. The clustering coefficient of a node $i$ is defined as $C_i=2n_i/k_i(k_i-1)$, where $n_i$ is the number of links connecting the $k_i$ neighbors of node $i$ to each other; equivalently, $C_i$ is the ratio between the number of triangles passing through two neighbors of $i$ and node $i$ itself, divided by the maximum possible number of such triangles. 
Random networks have low path lengths and low clustering coefficients, whereas regular lattices have long path lengths and are locally clustered. \citet{Watts1998a} have constructed an intermediate regime of ``small world'' networks, where the regular lattice has been perturbed by a small number of random links connecting distant parts of the network together. These networks, although not necessarily scale free, have short path lengths and high clustering coefficients, a property that was subsequently observed in metabolic and other biological networks as well \cite{Wagner2001a}.

A high clustering coefficient suggests the existence of densely connected groups of nodes within a network, which seems contradictory to the idea of scale invariance, in which there is no inherent group or cluster size;  \citet{Ravasz2002a}  addressed this problem by introducing hierarchical networks and providing a simple construction for synthetic hierarchical networks exhibiting both scale free and clustering behaviors. Although there is no unique scale for the clusters, clusters will appear at any scale one chooses to look at, and this is revealed by the scaling of clustering coefficient $C(k)$ with the node degree $k$, $C(k)\sim k^{-1}$, on both synthetic as well as natural metabolic networks of organisms from different domains of life \cite{Ravasz2002a}.   Another interesting property of some biological networks is an anticorrelation of node degree of connected nodes \cite{Maslov2002a}, which we can think of as a `dissociative' structure, in contrast, for example, with the associative character of social networks, where well connected people usually know one another.

As we look more finely at the structure of the graph representing a network, there is of course a much greater variety of things to look at. For example,  \citet{Spirin2003a} have  focused on high clustering coefficients as a starting point and devised algorithms to search for modules, or densely connected subgraphs within the yeast protein--protein interaction network. Although the problem has combinatorial complexity in general, the authors found about 50 modules (of 5-10 proteins in size, some of which were unknown at the time) that come in two types: the first represents dynamic functional units (e.g. signaling cascades), and the second protein complexes. A similar conclusion was reached by \citet{Han2004a}, after having analyzed  the interactome in combination with the temporal gene expression profiles and protein localization data; the authors argue that nodes of high degree can sit either at the centers of modules, which are simultaneously expressed (``party hubs''), or they can be involved in various pathways and modules at different times (``date hubs''). The former kind is at a lower level of organization, whereas the latter tie the network into one large connected component.

Focusing on even a smaller scale, \citet{Shen-Orr2002a} have explored motifs, or patterns of connectivity of small sets of nodes that are over represented in a given network compared to the randomized networks of the same degree distribution $P(k)$. In the transcriptional network of the bacterium \emph{E. coli}, three such motifs were found: feed forward loops (in which gene X regulates Y that regulates Z, but X directly regulates Z as well), single input modules (where gene X regulates a large number of other genes in the same way and usually autoregulates itself), and dense overlapping regulons (layers of overlapping interactions between genes and a group of transcription factors, much denser than in randomized networks). The motif approach has been extended to combined network of transcriptional regulation and 
protein--protein interactions \cite{Yeger-Lotem2004a} in yeast, as well as to other systems \cite{Milo2004a}.

At the risk of being overly pessimistic, we should conclude this section with a note of caution.  It would be attractive to think that a decade of work on network topology has resulted in a coherent picture, perhaps of the  following form: on the smallest scale, the nodes of biological networks are assembled into motifs, these in turn are linked into modules, and this continues in a hierarchical fashion until the entire network is scale free. As we will discuss again in the context of design principles, the notion of such discrete substructure---motifs and modules---is intuitively appealing, and some discussions suggest that it is essential either for the function or the evolution of networks.  On the other hand, the evidence for such structure usually is gathered with reference to some null model (e.g., a random network with the same $P(k)$), so we don't even have an absolute definition of these structures, much less a measure of their sufficiency as a characterization of the whole system; for  attempts at an absolute definition of modularity see \citet{Ziv2005b} and \citet{Hofman2007}.  Similarly, while it is appealing to think about scale free networks, the evidence for scaling almost always is confined to less than two decades, and in practice scaling often is not exact.  It is then not clear whether the idealization of scale invariance captures the essential structure in these systems.
\subsection{Boolean networks}
A straightforward  extension of the topological picture that also permits the study of network dynamics assumes that the entities at the nodes---for example, genes or signaling proteins---are  either `on' or `off' at each moment of time, so that for node $i$ the state at time $t$ is $\sigma_i(t)\in\left\{0,1\right\}$. Time is usually discretized, and an additional prescription is needed to implement the evolution of the system: $\sigma_i(t+1)=f_i(\left\{\sigma_\mu(t)\right\})$, where $f_i$ is a function the specifies how the states of the nodes $\mu$ that are the inputs to node $i$ in the interaction graph combine to determine the next state at node $i$. For instance, $f_A$ might be a Boolean function for gene $A$, which needs to have its activator gene $B$ resent and repressor gene $C$ absent, so that $\sigma_A(t+1)=\sigma_B(t) \wedge \bar{\sigma}_C(t)$. Alternatively, $f$ might be a function that sums the inputs at state $t$ with some weights, and then sets $\sigma_i = 1(0)$ if the result is above (below) a threshold, as in classical models of neural networks.

Boolean networks are amenable both to analytical treatment and to efficient simulation.  Early on \citet{Kauffman1969a} considered the family of random boolean networks.  In these models, each node is connected at random to $K$ other nodes on average, and it computes a random Boolean function of its inputs in which a fraction $\rho$ of the $2^K$ possible inputs combinations leads to $\sigma_i(t+1) =1$.  In the limit that the network is large, the dynamics are either regular
(settling into a stable fixed cycle) or chaotic, and these two phases are separated by a separatrix $2\rho(1-\rho)K=1$ in the phase space $(\rho,K)$.

\citet{Aldana2003a} have shown that for connectivities of $K\sim20$ that could reasonably be expected in e.g. transcriptional regulatory networks, the chaotic regime dominates the phase space. They point out, however, that if the network is scale free, there is no `typical' $K$ as the distribution $P(k)\sim k^{-\gamma}$ does not have a well-defined mean for $\gamma\leq 3$ and the phase transition criterion must be restated. It turns out, surprisingly, that regular behavior is possible for values of $\gamma$ between 2 and 2.5, observed in most biological networks, and this is exactly the region where the separatrix lies. Scale free architecture, at least for Boolean networks, seems to prevent chaos.

Several groups have used Boolean models to look at specific biological systems.  \citet{Thomas1973a} has established a theoretical framework in which current states of the genes (as well as the states in the immediate past) and the environmental inputs are represented by Boolean variables that evolve through the application of Boolean functions. This work has been continued by, for example, \citet{Sanchez2001a}, who analyzed the gap-gene system of the fruit fly \emph{Drosophila} by building a Boolean network that generates the correct levels of gene expression for 4 gap genes in response to  input levels of 3 maternal morphogens with spatially varying profiles stretched along the anterior-posterior axis of the fly embryo. Interestingly, to reproduce the observed results and correctly predict the known \emph{Drosophila} segmentation mutants, the authors had to introduce generalized Boolean variables that can take more than two states, and have identified the smallest necessary number of such states for each gene.

In a similar spirit, \citet{Li2004a}  studied the skeleton of the budding yeast cell cycle, composed of 11 nodes, and a thresholding update rule. They found  that the topology of this small network generates a robust sequence of transitions corresponding to known progression through yeast cell-cycle phases G1 (growth), S (DNA duplication), G2 (pause) and M (division), triggered by a known `cell-size checkpoint.' This progression is robust, in the sense that the correct trajectory is the biggest dynamical attractor of the system, with respect to various choices of update rules and parameters, small changes in network topology, and choice of triggering checkpoints.  

The usefulness of Boolean networks stems from the relative ease of implementation and simple parametrizations of network topology and dynamics, making them suitable for studying medium or large networks. In addition to simplifying the states at the nodes to two (or more) discrete levels, which is an assumption that has not been clearly explored, one should be cautious that the discrete and usually synchronous dynamics in time can induce unwanted artifacts.
\subsection{Probabilistic models}

Suppose one is able to observe simultaneously the activity levels of several proteins comprising a signaling network, or the expression levels of a set of genes belonging to the same regulatory module. Because they are part of a functional whole, the activity levels of the components will be correlated. Naively, one could build a network model by simply computing  pairwise correlation coefficients between pairs of nodes, and postulating an interaction, and therefore a link, between the two nodes whenever their correlation is above some threshold. However, in a test case where ${\rm A}\rightarrow {\rm B}\rightarrow{\rm C}$ (gene A induces B which induces C), one expects to see high positive correlation among all three elements, even though there is no (physical) interaction between A and C. Correlation  therefore is not equal to interaction or causation.  Constructing a network from the correlations in this naive way also does not lead to a generative model that would predict the probabilities for observing different states of the network as a whole.
Another approach is clearly needed; see \citet{Markowetz2007a} for a review.

In the simple case where the activity of a protein/gene $i$ can either be `on' ($\sigma_i=1$) or `off' ($\sigma_i=0$), the state of a network with $N$ nodes will be characterized by a binary word of $N$ bits, and because of interaction between nodes, not all these words will be equally likely. For example, if node A represses node B, then  combinations such as $1_A0_B\cdots$ or $0_A1_B\cdots$ will be more likely than $1_A1_B\cdots$. In the case of deterministic Boolean networks, having node A be `on' would imply that node B is `off' with certainty, but in probabilistic models it only means that there is a positive \emph{bias} for node B to be `off', quantified by  the probability that node B is `off' given that the state of node A is known. Having this additional probabilistic degree of freedom is advantageous, both because the network itself might be noisy, and because the experiment can induce errors in the signal readout, making the inference of deterministic rules from observed binary patterns an ill-posed problem. 

Once we agree to make a probabilistic model, the goal is to find the distribution over all network states, which we can also think of as the joint distribution of all the $N$ variable that live on the nodes of the network, 
$P(\sigma_1, \dots,\sigma_N|\mathcal{C})$, perhaps conditioned on some fixed set of environmental or experimental factors $\mathcal{C}$. The activities of the nodes, $\sigma_i$, can be binary, can take on a discrete set of states, or be continuous, depending on our prior knowledge about the system and experimental and numerical constraints. Even for a modest $N$, experiments of realistic scale  will not be enough to directly estimate the probability distribution, since even with binary variable the number of possible states, and hence the number of parameters required to specify the general probability distribution, grows as  $\sim 2^N$.  Progress thus depends in an essential way on simplifying assumptions.

Returning to the three gene example ${\rm A}\rightarrow {\rm B}\rightarrow{\rm C}$, we realize that C depends on A only through B, or in other words, $\rm C$ is \emph{conditionally independent} of $\rm A$ and hence no interaction should be assigned between nodes $\rm A$ and $\rm C$. 
Thus, the joint distribution of three variables can be factorized,
$$
P(\sigma_{\rm A} , \sigma_{\rm B} ,\sigma_{\rm C}) = P(\sigma_{\rm C} | \sigma_{\rm B}) P(\sigma_{\rm B} | \sigma_{\rm A}) P(\sigma_{\rm A}).
$$
One might hope that, even in a large network, these sorts of conditional independence relations could be used to simplify our model of the probability distribution.  In general this doesn't work, because of feedback loops which, in our simple example, would include the possibility that $\rm C$ affects the state of $\rm A$, either directly or through some more circuitous path.  Nonetheless one can try to make an approximation in which loops either are neglected or (more sensibly) taken into account in some sort of average way; in statistical mechanics, this approximation goes back at least to the work of \citet{Bethe1935}.

In the computer science and bioinformatics literature, the exploitation of Bethe--like approximations has come to be known as `Bayesian network modeling' \cite{Friedman2004a}.
In practice what this approach does is to search among possible network topologies, excluding loops, and then for fixed topology one uses the conditional probability relationships to factorize the probability distribution and fit the tables of conditional probabilities at each node that will best reproduce some set of data.  Networks with more links have more parameters, so one must introduce a tradeoff between the quality of the fit to the data and this increasing complexity.  In this framework there is thus an explicit simplification based on conditional independence, and an implicit simplification based on a preference for models with fewer links or sparse connectivity.   

The best known application of this approach to a biological network is the analysis of the MAPK signaling pathway in T cells from the human immune system \cite{Sachs2005a}.
The data for this analysis comes from experiments in which the phosophorylated states of 11 proteins in the pathway are sampled simultaneously by immunostaining \cite{Perez2002a}, with hundreds of cells sampled for each set of external conditions.    By combining experiments from multiple conditions, the Bayesian network analysis was able to find a network of interactions among the 11 proteins that has high overlap with those known to occur experimentally. 

A very different approach to simplification of probabilistic models is based on the maximum entropy principle \cite{Jaynes1957}.  In this approach one view a set of experiments as providing an estimate of some set of correlations, for example the $\sim N^2$ correlations among all pairs of elements in the network.  One then tries to construct a probability distribution which matches these correlations but otherwise has as little structure---as much entropy---as possible.  We recall that the Boltzmann distribution for systems in thermal equilibrium can be derived as the distribution which has maximum entropy consistent with a given average energy, and maximum entropy modeling generalizes this to take account of other average properties.  In fact one can construct a hierarchy of maximum entropy distributions which are consistent with higher and higher orders of correlation \cite{Schneidman2003b}.  Maximum entropy models for binary variables that are consistent with pairwise correlations are exactly the Ising models of statistical physics, which opens a wealth of analytic tools and intuition about collective behavior in these systems.

In the context of biological networks (broadly construed), recent work has shown that maximum entropy models consistent with pairwise correlations are surprisingly successful at describing the patterns of activity among populations of neurons in the vertebrate retina as it responds to natural movies \cite{Schneidman2006a, Tkacik2006a}. Similar results are obtained for very different retinas under different conditions \cite{Shlens2006a}, and these successes have touched a flurry of interest in the analysis of neural populations more generally.    The connection to the Ising model has a special resonance in the context of neural networks, where the collective behavior of the Ising model has been used for some time as a prototype for thinking about the dynamics of computation and memory storage \cite{Hopfield1982a}; in the maximum entropy approach the Ising model emerges directly as the least structured model consistent with the experimentally measured patterns of correlation among pairs of cells.  A particularly striking result of this analysis is that the Ising models which emerge seem to be poised near a critical point \cite{Tkacik2006a}.   Returning to cell biology, the maximum entropy approach has also been used to analyze patterns of gene expression in yeast \cite{Lezon2006} as well as to revisit the MAPK cascade \cite{Tkacik2007d}.   

\subsection{Dynamical systems}
If the information about a biological system is detailed enough to encompass all relevant interacting molecules along with the associated reactions and estimated reaction rates, and the molecular noise is expected to play a negligible role, it is possible to describe the system with rate equations of chemical kinetics. An obvious benefit is the immediate availability of mathematical tools, such as steady state and stability analyses, insight provided by nonlinear dynamics and chaos theory, well developed numerical algorithms for integration in time and convenient visualization with phase portraits or bifurcation diagrams. Moreover, analytical approximations can be often exploited productively when warranted by some prior knowledge, for example, in separately treating `fast' and `slow' reactions. In practice, however, reaction rates and other important parameters are often unknown or known only up to order-of-magnitude estimations; in this case the problem usually reduces to the identification of phase space regions where the behavior of the system is qualitatively the same, for example, regions where the system exhibits limit-cycle oscillations, bistability, convergence into a single steady state etc.; see \citet{Tyson2001a} for a review.
Despite the difficulties,  deterministic chemical kinetic models have been very powerful tools  in analyzing specific network motifs or regulatory elements, as in the protein signaling circuits that achieve perfect adaptation, homeostatsis, switching and so on described by \citet{Tyson2003a}, and more generally in the analysis of transcriptional regulatory networks as reviewed by  \citet{Hasty2001a}.

In the world of bacteria, some of the first detailed computer simulation of the chemotaxis module of \emph{Escherichia coli} were undertaken by \citet{Bray1993a}. The signaling cascade from the Tar receptor at the cell surface to the modifications in the phosphorylation state of the molecular motor were captured by Michaelis-Menten kinetic reactions (and equilibrium binding conditions for the receptor), and the system of equations was numerically integrated in time. While slow adaptation kinetics was not studied by in this first effort, the model nevertheless qualitatively reproduces about 80 percent of examined chemotactic protein deletion and overexpression  mutants, although the extreme sensitivity of the system remained unexplained. 

In eukaryotes, \citet{Novak1997a} have, for instance, constructed an extensive model of cell cycle control in fission yeast.  Despite its complexity ($\sim\!10$ proteins and $\sim\!30$ rate constants), Novak and colleagues have provided an interpretation of the system in terms of three interlocking modules that regulate the transitions from G1 (growth) into S (DNA synthesis) phase, from G2 into M (division) phase, and the exit from mitosis, respectively. The modules are coupled through cdc2/cdc13 protein complex and the system is driven by the interaction with the cell size signal (proportional to the number of ribosomes per nucleus). At small size, the control circuit can only support one stable attractor, which is the state with low cdc2 activity corresponding to G1 phase. As the cell grows, new stable state appears and the system makes an irreversible transition into S/G2 at a bifurcation point, and, at an even larger size, the mitotic module becomes unstable and executes limit cycles in cdc2-cdc13 activity until the M phase is completed and the cell returns to its initial size. The basic idea is that the cell, driven by the the size readout, progresses through robust cell states created by bistability in the three modules comprising the cell cycle control -- in this way, once it commits to a transition from G2 state into M, small fluctuations will not flip it back into G2. The mathematical model has in this case successfully predicted the behaviors of a number of cell cycle mutants and recapitulated experimental observations collected during 1970s and 1980s by Nurse and collaborators
\cite{Nurse}.

The circadian clock is a naturally occurring transcriptional module that is particularly amenable to dynamical systems modeling. \citet{Leloup2003a} have created a mathematical model of a mammalian clock (with $\sim\!20$ rate equations) that exhibits autonomous sustained oscillations over a sizable range of parameter values, and reproduces the entrainment of the oscillations to the light--dark cycles through light-induced gene expression. The basic mechanism that enables the cyclic behavior is negative feedback transcriptional control, although the actual circuit contains at least two coupled oscillators. Studying circadian clock in mammals, the fruit fly \emph{Drosophila} or \emph{Neurospora} is attractive because of the possibility of connecting a sizable catalogue of physiological disorders in circadian rhythms to malfunctions in the clock circuit and direct experimentation with light-dark stimuli \cite{Young2001a}.  Recent experiments indicate that at least in cyanobacteria the circdian clock can be reconstituted from a surprisingly small set of biochemical reactions, without transcription or translation \cite{Tomita2005,Nakajima2005}, and this opens possibilities for even simpler and highly predictive dynamical models \cite{Rust2007}.

Dynamical modeling has in addition been applied to many smaller systems. For example, the construction of a synthetic toggle switch \cite{Gardner2000a}, and the `repressilator' -- oscillating network of three mutually repressing genes \cite{Elowitz2000a} -- are examples where mathematical analysis has stimulated the design of synthetic circuits.  A successful reaction-diffusion model of how localization and complex formation of Min proteins can lead to spatial limit cycle oscillations (used by  \emph{Escherichia coli}  to find its division site)  was constructed by \citet{Huang2003a}. 
It remains a challenge, nevertheless, to  navigate in the space of parameters as it becomes ever larger for bigger networks, to correctly account for localization and count various forms of protein modifications, especially when the signaling networks also couple to transcriptional regulation, and  to find a proper balance between models that capture all known reactions and interactions and phenomenological models that include coarse-grained variables.
\subsection{Stochastic dynamics}
Stochastic dynamics is in principle the most detailed level of system description. Here, the (integer) count of every molecular species is tracked and reactions are drawn at random with appropriate probabilities per unit time (proportional to their respective reaction rates) and executed to update the current tally of  molecular counts. An algorithm implementing this prescription, called  the stochastic simulation algorithm or SSA, was devised by \citet{Gillespie1977a}; see \citet{Gillespie2007a} for a review of SSA and a discussion of related methods. Although slow, this approach simulating chemical reactions can be made exact.  In general, when all molecules are present in large numbers and continuous, well-mixed concentrations are good approximations, the (deterministic) rate dynamics equations and stochastic simulation give the same results; however, when molecular counts are low and, consequently, the stochasticity in reaction timing and ordering becomes important,  the rate dynamics breaks down and SSA needs to be used. In biological networks and specifically in transcriptional regulation, a gene and its promoter region are only present in one (or perhaps a few) copies, while transcription factors that regulate it can also be at nanomolar concentrations (i.e. from a few to a few hundred molecules per nucleus), making stochastic effects possibly very important \cite{McAdams1997a, McAdams1999a}.

One of the pioneering studies of the role of noise in a biological system was a simulation of the phage $\lambda$ lysis-lysogeny switch by \citet{Arkin1998a}. The lifecycle of the phage is determined by the concentrations of two transcription factors, \emph{cI} (lambda repressor) and \emph{cro}, that compete for binding to the same operator on the DNA. If \emph{cI} is prevalent, the phage DNA is  integrated into the host's genome and no phage genes except for \emph{cI} are expressed (the lysogenic state); if \emph{cro} is dominant, the phage is in lytic state, using cell's DNA replication machinery to produce more phages and ultimately lyse the host cell \cite{Ptashne2004a}. The switch is bistable and the fate of the phage depends on the temporal and random pattern of gene expression of two mutually antagonistic transcription factors, although the balance can be shifted by subjecting the host cell to stress and thus flipping the toggle into lytic phase. The stochastic simulation correctly reproduces the experimentally observed fraction of lysogenic phages as a function of multiplicity-of-infection. An extension of SSA to spatially extended models is possible.

Although the simulations are exact, they are computationally intensive and do not offer any analytical insight into the behavior of the solutions. As a result, various theoretical techniques have been developed for studying the effects of stochasticity in biological networks. These are often operating in a regime where the deterministic chemical kinetics is a good approximation, and noise (i.e. fluctuation of concentrations around the mean) is added into the system of differential equations as a perturbation; these Langevin methods have  been useful for the study of noise propagation in regulatory networks \cite{vanKampen2007a, Thattai2001a, Paulsson2004a}.
The analysis of stochastic dynamics is especially interesting in the context of design principles which consider the reliability of network function, to which we return below.
\section{Network properties and operating principles}
\subsection{Modularity}
Biological networks are said to be modular, although the term has several related but nevertheless distinct meanings. Their common denominator is the idea that there exist a partitioning of the network nodes into groups, or modules, that are largely independent of each other and perform separate or autonomous functions. Independence can be achieved through spatial isolation of  the module's processes or by chemical specificity of its components. The ability to extract the module from the cell and reconstitute it \emph{in vitro}, or transplant it to another type of cell is a powerful argument for  the existence of modularity \cite{Hartwell1999a}. In the absence of such strong and laborious experimental verifications, however, measures of modularity that depend on a particular network model are frequently used.

In topological networks, the focus is on the module's independence: nodes within a module are densely connected to each other, while inter-modular links are sparse \cite{Ravasz2002a, Spirin2003a, Han2004a} and the tendency to cluster is measured by high clustering coefficients. As a caveat to this view note that despite their sparseness the inter-module links could represent strong dynamical couplings. Modular architecture has been studied in Boolean networks by \citet{Kashtan2005a}, who have shown that modularity can evolve by mutation and selection in a time-varying fitness landscape where changeable goals decompose into a set of fixed subproblems. In the example studied they computationally evolve networks implementing several Boolean formulae and observe the appearance of a module -- a circuit of logical gates implementing a particular Boolean operator (like XOR) in a reusable way. 
This work makes clear that modularity in networks is plausibly connected to modularity in the kinds of problems that these networks were selected to solve, but we really know relatively little about the formal structure of these problems.

There are also ways of inferring a form of modularity directly without assuming any particular network model. Clustering tools partition genes into co-expressed groups, or clusters, that are often identified with particular modules \cite{Eisen1998a, Segal2003a,Slonim2005a}. \citet{Ihmels2002a} have noted that each node can belong to more than one module depending on the biological state of the cell, or the context, and have correspondingly reexamined the clustering problem. \citet{Elemento2007a} have recently presented a general information theoretic approach to inferring regulatory modules and the associated transcription factor binding sites from various kinds of high-throughput data.
While clustering methods have been widely applied in the exploration of gene expression, it should be emphasized that merely finding clusters does not by itself provide evidence for modularity.  As noted above, the whole discussion would be much more satisfying if we had independent definitions of modularity and, we might add, clearly stated alternative hypotheses about the structure and dynamics of these networks.

Focusing on the functional aspect of the module, we often observe that the majority of the components of a system (for instance, a set of promoter sites or a set of genes regulating motility in bacteria) are conserved together across species. These observations support the hypothesis that the conserved components are part of a very tightly coupled sub-network which we might identify as a module. Bioinformatic tools can then use the combined sequence and expression data to give predictions about modules, as reviewed by  \citet{Siggia2005a}.  Purely phylogenetic approaches that infer module components based on inter-species comparisons have also been productive and can extract candidate modules based only on phylogenetic footprinting, that is, studying the  presence or absence of homologous genes across organisms and correlating their presence with hand annotated phenotypic traits  \citep{Slonim2006a}.
\subsection{Robustness}
\emph{Robustness} refers to a property of the biological network such that some aspect of its function is not sensitive to perturbations of network parameters, environmental variables (e.g. temperature), or initial state;  see \citet{deVisser2003a} for a review of robustness from an evolutionary perspective and \citet{Goulian2004a} for mechanisms of robustness in bacterial circuits. Robustness encompasses two very different ideas.  One idea has to do with a general principle about the nature of explanation in the quantitative sciences:  qualitatively striking facts should not depend on the fine tuning of parameters, because such a scenario just shifts the problem to understanding why the parameters are tuned as they are.   The second idea is more intrinsic to the function of the system, and entails the hypothesis that cells cannot rely on precisely reproducible parameters or conditions and must nonetheless function reliably and reproducibly.

Robustness has been studied extensively in the chemotactic system of the bacterium \emph{Escherichia  coli}. The systematic bias to swim towards chemoattractants and away from repellents can only be sustained if the bacterium is sensitive to the spatial gradients of the concentration and not to its absolute levels. This discriminative ability is ensured by the mechanism of  perfect adaptation, with which the proportion of bacterial straight runs and tumbles (random changes in direction) always returns to the same value in the absence of gradients \cite{Block1983a}. Naively, however, the ability to adapt perfectly seems to be sensitive to the amounts of intracellular signaling proteins, which can be tuned only approximately by means of transcriptional regulation.  \citet{Barkai1997a} argued that there is integral feedback control in the chemotactic circuit which makes it robust against changes in these parameters, and \citet{Alon1999a} showed experimentally that precision of adaptation truly stays robust, while other properties of the systems (such as the time to adapt and the steady state) show marked variations with intracellular signaling protein concentrations.

One seemingly clear example of robust biological function is embryonic development.  We know that the spatial structure of the fully developed organism follow a `blueprint' laid out early in development as a spatial pattern of gene expression levels.  \citet{vonDassow2000a} studied one part of this process in the fruit fly {\em Drosophila}, the `segment polarity network' that generates striped patterns of expression.  
They considered  a dynamical system based on the wiring diagram of interactions among a smal group of genes and signaling molecules,  with $\sim\!50$ associated constants parametrizing production and degradation rates, saturation response and diffusion, and searched the parameter space for solutions that reproduce the known striped patterns.  They found that, with their initial guess at network topology, such solutions do not exist, but adding a particular link -- biologically motivated though unconfirmed at the time -- allowed them to find solutions by random sampling of parameter space. Although they presented no rigorous measure for the volume of parameter space in which correct solutions exist, it seems that a wide variety of parameter choices and initial conditions indeed produce striped expression patterns, and this was taken to be a signature of robustness.

Robustness in dynamical models is the ability of the biological network to sustain its trajectory through state space despite parameter or state perturbations. In circadian clocks the oscillations have to be robust against both molecular noise inherent in transcriptional regulation, examined in stochastic simulations by \citet{Gonze2002a}, as well as variation in rate parameters \cite{Stelling2004a}; in the latter work the authors introduce integral robustness measures along the trajectory in state space and argue that the clock network architecture tends to concentrate the fragility to perturbations into parameters that are global to the cell (maximum overall translation and protein degradation rates) while increasing the robustness to processes specific to the circadian oscillator. 
As was mentioned earlier, robustness to state perturbations was demonstrated by \citet{Li2004a} in the threshold binary network model of the yeast cell cycle, and examined in scale-free random Boolean networks by \citet{Aldana2003a}.

As with modularity, robustness has been somewhat resistant to rigorous definitions.  Importantly, robustness has always been used as a relational concept:  function $X$ is robust to variations in $Y$.  An alternative to robustness is for the organism to exert precise control over $Y$, perhaps even using $X$ as a feedback signal.  This seems to be how neurons stabilize a functional mix of different ion channels \cite{Marder2006a}, following the original theoretical suggestion of \citet{LeMasson1993a}.  Pattern formation during embryonic development in {\em Drosophila} begins with spatial gradients of transcription factors, such as Bicoid, which are established by maternal gene expression, and it has been assumed that variations in these expression levels are inevitable, requiring some robust readout mechanism.  Recent measurements of Bicoid in live embryos, however, demonstrate that the absolute concentrations are actually reproducible from embryo to embryo with $\sim 10\%$ precision \cite{Gregor2007a}.  While there remain many open questions, these results suggest that organisms may be able to exert surprisingly exact control over critical parameters, rather than having compensation schemes for initially sloppy mechanisms.  The example of ion channels alerts us to the possibility that cells may even `know' which combinations of parameters are critical, so that variations in a multidimensional parameter space are large, but confined to a low dimensional manifold.

\subsection{Noise}
A dynamical system  with constant reaction rates, starting repeatedly from the same initial condition in a stable environment, always follows a deterministic time evolution. When the concentrations of the reacting species are low enough, however, the description in terms of time (and possibly space) dependent concentration breaks down, and the stochasticity in reactions, driven by random encounters between individual molecules, becomes important: on repeated trials from the same initial conditions, the system will trace out different trajectories in the state space. As has been pointed out in the section on stochastic dynamics, biological networks in this regime need to be simulated with the Gillespie algorithm \cite{Gillespie1977a}, or analyzed within approximate schemes that treat noise as perturbation of deterministic dynamics. Recent experimental developments have made it possible to observe this noise directly, spurring new research in the field.  
Noise in biological networks fundamentally limits the organism's ability to sense, process and respond to environmental and internal signals, suggesting that analysis of noise is a crucial component in any attempt to understand the design of these networks.  This line of reasoning is well developed in the context of neural function \cite{Bialek1987a}, and we draw attention in particular to work on the ability of the visual system to count single photons, which depends upon the precision of the G-protein mediated signaling cascade in photoreceptors; see, for example, \citet{Ramanathan2005a}.

Because transcriptional regulation inherently deals with molecules, such as DNA and transcription factors,  that are present at low copy numbers, most noise studies were carried out on transcriptional regulatory elements. The availability of fluorescent proteins and their fusions to wild type proteins have been the crucial tools, enabling researchers to image the cells expressing these probes in a controllable manner, and track their number in time and across the population of cells. \citet{Elowitz2002a} pioneered the idea of observing the output of two identical regulatory elements  driving the expression of two fluorescent proteins of different colors, regulated by a common input in a single \emph{Escherichia coli} cell. In this `two-color experiment,' the correlated fluctuations in both colors must be due to the \emph{extrinsic} fluctuations in the common factors that influence the production of both proteins, such as overall RNA polymerase or transcription factor levels; on the other hand, the remaining, uncorrelated fluctuation is due to the \emph{intrinsic} stochasticity in the transcription of the gene and translation of the messenger RNA into the fluorescent protein from each of the two promoters \cite{Swain2002a}. \citet{Ozbudak2002a} have studied the contributions of stochasticity in transcription and translation to the total noise in gene expression in prokaryotes, while \citet{Pedraza2005a} and \citet{Hooshangi2005a} have looked at the propagation of noise from transcription factors to their targets in synthetic multi-gene cascades. \citet{Rosenfeld2005a} have used the statistics of binomial partitioning of proteins during the  division of \emph{Escherichia coli} to convert their fluorescence measurements into the corresponding absolute protein concentrations, and also were able to observe the dynamics of these fluctuations, characterizing the correlation times of both intrinsic and extrinsic noise.

Theoretical work has primarily been concerned with disentangling and quantifying the contributions of different steps in transcriptional regulation and gene expression to the total noise in the regulated gene \cite{Thattai2001a, Paulsson2004a, Swain2004a}, often by looking for signatures of various noise sources in the behavior of the measured noise as a function of the mean expression level of a gene. For many of the examples studied in prokaryotes, noise seemed to be primarily attributable to the production of proteins in bursts from single messenger RNA molecules, and to pulsatile and random activation of genes and therefore bursty translation into mRNA \cite{Golding2005a}. In  yeast \cite{Blake2003a,Raser2005a} and in mammalian cells \cite{Raj2006a} such stochastic synthesis of mRNA was modeled and observed as well. Simple scaling of noise with the mean was observed in $\sim\!40$ yeast proteins under different conditions by \citet{Bar-Even2006a} and interpreted as originating in variability in mRNA copy numbers or gene activation.

\citet{Bialek2005a} have demonstrated theoretically that at low concentrations of transcriptional regulator, there should be a lower bound on the noise set by the basic physics of diffusion  of transcription factor molecules to the DNA binding sites.  This limit is independent of
(possibly complex, and usually unknown) molecular details of the binding process; as an example, cooperativity enhances the `sensitivity' to small changes in concentration, but doesn't lower the physical limit to noise performance \cite{Bialek2006a}. This randomness in diffusive flux of factors to their `detectors' on the DNA must ultimately limit the precision and reliability of transcriptional regulation, much like the randomness in diffusion of chemoattractants to the detectors on the surface of \emph{Escherichia coli}    limits its chemotactic performance \cite{Berg1977a}. Interestingly, one dimensional diffusion of transcription factors along the DNA can have a big impact on the speed with which TFs find their targets, but the change in noise performance that one might expect to accompany these kinetic changes is largely compensated by the extended correlation structure of one dimensional diffusion \cite{Tkacik2007e}.
Recent measurements of the regulation of the {\em hunchback} gene by Bicoid during early fruit fly development by \citet{Gregor2007a} have provided evidence for the dominant role of such input noise, which coexists with previously studied output noise in production of mRNA and protein.  These results raise the possibility that initial decisions in embryonic development are made with a precision limited by fundamental physical principles.
\subsection{Dynamics, attractors, stability and large fluctuations}
The behavior of a dynamical system as the time tends to infinity, in response to a particular input, is interesting regardless of the nature of the network model. Both discrete and continuous, or deterministic and noisy, systems can settle into a number of fixed points, exhibit limit-cycle oscillations, or execute chaotic dynamics. In biological networks it is important to ask whether these qualitatively different outcomes correspond to distinct phenotypes or behaviors. If so, then a specific stable gene expression profile in a network of developmental genes, for example, encodes that cell's developmental fate, as the amount of lambda repressor encodes the state of lysis vs lysogeny switch in the phage. The history of the system that led to the establishment of a specific steady state would not matter as long as the system persisted in the same attractor: the dynamics could be regarded as a `computation' leading to the final result, the identity of the attractor, with the activities of genes in this steady state in turn driving the downstream pathways and other modules; see \citet{Kauffman1969a} for genetic networks and \citet{Hopfield1982a} for similar ideas in neural networks for associative memory. Alternatively, such partitioning into transient dynamics and `meaningful' steady states might not be possible: the system must be analyzed as a whole while it moves in state space, and parts of it do not separately and sequentially settle into their attactors. 

It seems, for example, that qualitative behavior of the cell cycle can be understood by progression through well-defined states or checkpoints: after transients die away, the cell cycle proteins are in a `consistent' state that regulates division or growth related activities, so long as the conditions do not warrant a new transition into the next state \cite{Nasmyth1996a,Chen2000a}. In the fruit fly \emph{Drosophila} development it has been suggested that combined processes of diffusion and degradation first establish steady-state spatial profiles of maternal morphogens along the major axis of the embryo, after which this stable `coordinate system' is read out by gap and other downstream genes to generate the body segments. Recent measurements by \citet{Gregor2007b} have shown that there is a rich dynamics in the Bicoid morphogen concentration, prompting \citet{Bergmann2007a} to hypothesize that perhaps downstream genes read out and respond to morphogens even before the steady state has been reached. On another note, an interesting excitable motif, called the ``feedback resistor,'' has been found in HIV Tat system -- instead of having a bistable switch like the lambda phage, HIV (which lacks negative feedback capability) implements a circuit with a single stable `off' lysogenic state, that is perturbed in a pulse of transactivation when the virus attacks. The pulse probably triggers a threshold-crossing process that drives downstream events, and subsequently decays away; the feedback resistor is thus again an example of a dynamic, as opposed to the steady-state, readout \cite{Weinberger2007a}.   Excitable dynamics are of course at the heart of the action potential in neurons, which results from the coupled dynamics of ion channel proteins, and related dynamical ideas are now emerging other cellular networks \cite{Suel2006}.

If attractors of the dynamical system correspond to distinct biological states of the organism, it is important to examine their stability against noise-induced spontaneous flipping. Bistable elements are akin to the `flip-flop' switches in computer chips -- they form the basis of cellular (epigenetic) memory. While this mechanism for remembering the past is not unique -- for example, a very slow, but not bistable, dynamics will also retain `memory' of the initial condition through protein levels that persist on a generation time scale \cite{Sigal2006a}, it has the potential to be the most stable mechanism. The naturally occurring bistable switch of the lambda phage was studied using stochastic simulation by \citet{Arkin1998a}, and a synthetic toggle switch was constructed in \emph{Escherichia coli} by \citet{Gardner2000a}. Theoretical studies of systems where large fluctuations are important are generally difficult and restricted to simple regulatory elements, but \citet{Bialek2001a} has shown that a bistable switch can be created with as few as tens molecules  yet remain stable for years.
A full understanding of such stochastic switching brings in powerful methods from statistical physics and field theory \cite{Sasai2003a, Walczak2005a, Roma2005}, ultimately with the hope of connecting to quantitative experiments \cite{Acar2005}.
\subsection{Optimization principles}
If the function of a pathway or a network module can be quantified by a scalar measure, it is possible to explore the space of networks that perform the given function optimally. An example already given was that of maximizing the growth rate of the bacterium \emph{Escherichia coli}, subject to the constraints imposed by the known metabolic reactions of the cell; the resulting optimal joint usage of oxygen and food could be compared to the experiments \cite{Ibarra2002a}. If enough constraints exist for the problem to be well posed, and there is sufficient reason to believe that evolution drove the organism towards optimal behavior, optimization principles allow us to both tune the otherwise unknown parameters to achieve the maximum, and also to compare the wild type and optimal performances. 

\citet{Dekel2005a} have performed the cost/benefit analysis of expressing \emph{lac} operon in bacteria. On one hand \emph{lac} genes allow  \emph{Escherichia coli} to digest lactose, but on the other there is the incurred metabolic cost to the cell for expressing them. That the cost is not negligible to the bacterium is demonstrated best by the fact that it  shuts off the operon if no lactose is present in the environment. The cost terms are measured by inducing the \emph{lac} operon with changeable amount of IPTG that provides no energy in return; the benefit is measured by fully inducing \emph{lac} with IPTG and supplying variable amounts of lactose; both cost and benefit are in turn expressed as the change in the growth rate compared to the wild-type grown at fixed conditions. Optimal levels of \emph{lac} expression were then predicted as a function of lactose concentration and bacteria were evolved for several hundred generations to verify that evolved organisms lie close to the predicted optimum. 

\citet{Zaslaver2004a} have considered a cascade of amino-acid biosynthesis reactions in  \emph{Escherichia coli}, catalyzed by their corresponding enzymes. They have then optimized the parameters  of the model that describes the regulation of enzyme gene expression, such that the total metabolic cost for enzyme production was balanced against the benefit of achieving a desired metabolic flux through the biosynthesis pathway. The resulting optimal on-times and promoter activities for the enzymes were compared to the measured activities of amino-acid biosynthesis promoters exposed to different amino-acids in the medium. The authors conclude that the bacterium implements a `just-in-time' transcription program, with enzymes catalyzing initial steps in the pathway being produced from strong and low-latency promoters.

In signal transduction networks the definition of an objective function to be maximized is somewhat more tricky. The ability of the cell to sense its environment and make decisions, for instance about which genes to up- or down-regulate, is limited by several factors: scarcity of signals coming from the environment, perhaps because of the limited time that can be dedicated to data collection; noise inherent in the signaling network that degrades the quality of the detected signal; (sub-)optimality of the decision strategy; and noise in the effector systems at the output. A first idea would be to postulate that networks are designed to lower the noise, and intuitively the ubiquity of mechanisms such as negative feedback \cite{Becskei2000a, Goulian2004a} is consistent with such an objective. There are various definitions for noise, however, which in addition are generally a function of the input, raising serious issues about how to formulate a principled optimization criterion.

When we think about energy flow in biological systems, there is no doubt that our thinking must at least be consistent with thermodynamics.  More strongly, thermodynamics provides us with notions of efficiency that place the performance of biological systems on an absolute scale, and in many cases this performance really is quite impressive. 
In contrast, most discussions of  information in biological systems leave ``information'' as a colloquial term, making no reference to the formal apparatus of information theory as developed by Shannon and others more than fifty years ago \cite{Shannon1948}.
Although many aspects of information theory that are especially important for modern technology (e.g., sophisticated error--correcting codes) have no obvious connection to biology, there is something at the core of information theory that is vital:  Shannon proved that if we want to quantify the intuitive concept that ``$x$ provides information about $y$,'' then there is only one way to do this that is guaranteed to work under all conditions and to obey simple intuitive criteria such as the additivity of independent information.  This unique measure of ``information'' is Shannon's mutual information.  Further, there are theorems in information theory which, in parallel to results in thermodynamics, provide us with limits to what is possible and with notions of efficiency.  

There is a long history of using information theoretic ideas to analyze the flow of information in the nervous system, including the idea that aspects of the brain's coding strategies might be chosen to optimize the efficiency of coding, and these theoretical ideas have led directly to interesting experiments.  The use of information to think about cellular signaling and its possible optimization is more recent \cite{Ziv2006a,Tkacik2007a}.  An important aspect of optimizing information flow is that the input/output relations of signaling devices must be matched to the distribution of inputs, and recent measurements on the control of {\em hunchback} by Bicoid in the early fruit fly embryo \cite{Gregor2007a} seem remarkably consistent with the (parameter free) predictions from these matching relations \cite{Tkacik2007b}.

In the context of neuroscience there is a long tradition of forcing the complex dynamics of signal processing into a setting where the subject needs to decide between a small set of alternatives; in this limit there is a well developed theory of optimal Bayesian decision making, which uses prior knowledge of the possible signals to help overcome noise intrinsic to the signaling system;
\citet{Libby2007a} have recently applied this approach to the \emph{lac} operon in \emph{Escherichia coli}. The regulatory element is viewed as an inference module that has to `decide,' by choosing its induction level, if the environmental lactose concentration is high or low. If the bacterium detects a momentarily high sugar concentration, it has to discriminate between two situations: either the environment really is at low overall concentration but there has been a large fluctuation; or the environment has switched to a high concentration mode. The authors examine how plausible regulatory element architectures (e.g. activator vs repressor, cooperative binding etc.) yield different discrimination performance. Intrinsic noise in the \emph{lac} system can additionally complicate such decision making, but can be included into the theoretical Bayesian framework.

The question of whether biological systems are optimal in any precise mathematical sense is likely to remain controversial for some time.  Currently opinions are stronger than the data, with some investigators using `optimized' rather loosely and others convinced that what we see today is only a historical accident, not organizable around such lofty principles.   We emphasize, however, that attempts to formulate optimization principles require us to articulate clearly what we mean by ``function'' in each context, and this is an important exercise.  Exploration of optimization principles also exposes new questions, such as the nature of the distribution of inputs to signaling systems, that one might not have thought to ask otherwise.  Many of these questions remain as challenges for a new generation of experiments.
\subsection{Evolvability and designability} 
\citet{Kirschner1998a} define \emph{evolvability} as an organism's capacity to generate heritable phenotypic variation. This capacity may have two components: first, to reduce the lethality of mutations, and second, to reduce the number of mutations needed to produce phenotypically novel traits.  The systematic study of evolvability is hard because the genotype-to-phenotype map is highly non-trivial, but there have been some qualitative observations relevant to biological networks. Emergence of \emph{weak linkage} of processes, such as the co-dependence of transcription factors and their DNA binding sites in metazoan transcriptional regulation, is one example. Metazoan regulation seems to depend on combinatorial control by many transcription factors with weak DNA-binding specificities and the corresponding binding sites (called cis-regulatory modules) can be dispersed and extended on the DNA. This is in stark contrast to the strong linkage between the factors and the DNA in prokaryotic regulation or in metabolism, energy transfer or macromolecular assembly, where steric and complementarity requirements for interacting molecules are high. In protein signaling networks, strongly  conserved but flexible proteins, like calmodulin, can bind weakly to many other proteins, with small mutations in their sequence probably affecting such binding and making the establishment of new regulatory links possible and perhaps easy.

Some of the most detailed attempts to follow the evolution of network function have been by Francois and coworkers \cite{Francois2004a,Francois2007a}.  In their initial work they showed how 
 simple functional circuits, performing logical operations or implementing bistable or oscillatory behavior, can be reliably created by a mutational process with selection by an appropriate fitness function.   More recently they have considered fitness functions which favor spatial structure in patterns of gene expression, and shown how the networks that emerge from dynamics in this fitness landscape recapitulate the outlines of the segmentation networks known to be operating during embryonic development.

Instead of asking if there \emph{exists} a network of nodes such that they perform a given computation, and if it can be found by mutation and selection as in the examples above, one can ask how many network topologies perform a given computation. In other words, one is asking whether there is only one (fine tuned?) or many topologies or solutions to a given problem. The question of how many network topologies, proxies for different genotypes, produce the same dynamics, a proxy for phenotype, is a question of designability, a concept originally proposed to study the properties of amino-acid sequences comprising functional proteins, but applicable also to biological regulatory networks \cite{Nochomovitz2006a}. The authors examine three- and four-node binary networks with threshold updating rule and show that all networks with the shared phenotype have  a common `core' set of connections, but can differ in the variable part, similar to protein folding where the essential set of residues is necessary for the fold, with numerous variations in the nonessential part.
\section{Future prospects}
The study of biological networks is at an early stage, both on the theoretical as well as on the experimental side. Although high-throughput experiments are generating large datasets, these can suffer from serious biases, lack of temporal or spatial detail, and limited access to the component  parts of the interacting system. On a theoretical front, general analytical insights that would link dynamics with network topology are few, although for specific systems with known topology computer simulation can be of great assistance. There can be confusion about which aspects of the dynamical model have biological significance and interpretation, and which aspects are just `temporary variables' and the `envelope' of the proverbial back-of-the-envelope calculations that cells use to perform their biological computations on; which parts of the trajectory are functionally constrained and which ones could fluctuate considerably with no ill-effects; how much noise is tolerable in the nodes of the network and what is its correlation structure; or how the unobserved, or `hidden', nodes (or their modification/activity states) influence the network dynamics.

Despite these caveats, cellular networks have some advantages over biological systems of comparable complexity, such as neural networks. Due to technological developments, we are considerably closer to the complete census of the interacting molecules in a cell than we are generally to the picture of connectivity of the neural tissue. Components of the regulatory networks are simpler than neurons, which are capable of a range of complicated behaviors on different timescales. Modules and pathways often comprise smaller number of interacting elements than in neural networks, making it possible to design small but interesting synthetic circuits. Last but not least, sequence and homology can provide strong insights or be powerful tools for network inference in their own right.

Those of us who come from the traditionally quantitative sciences, such as physics, were raised with experiments in which crucial elements are isolated and controlled.  In biological systems, attempts at such isolation may break the regulatory mechanisms that are essential for normal operation of the system, leaving us with a system which is fact more variable and less controlled than we would have if we faced the full complexity of the organism.  It is only recently that we have seen the development of experimental techniques that allow fully quantitative, real time measurements of the molecular events inside individual cells, and the theoretical framework into which such measurements will be fit still is being constructed.  The range of theoretical approaches being explored is diverse, and it behooves us to search for those approaches which have the chance to organize our understanding of many different systems rather than being satisfied with models of particular systems.  Again, there is a balance between the search for generality and the need to connect with experiments on specific networks.  We have tried to give some examples of all these developments, hopefully conveying the correct combination of enthusiasm and skepticism.

\begin{acknowledgments}
We thank our colleagues and collaborators who have helped us learn about these issues:  MJ Berry, CG Callan, T Gregor, JB Kinney, P Mehta, SE Palmer, E Schneidman, JJ Hopfield, T Mora, S Setayeshgar, N Slonim, GJ Stephens, DW Tank, N Tishby, A Walczak, EF Wieschaus,  CH Wiggins and NS Wingreen.  Our work was supported in part by NIH grants P50 GM071508 and R01 GM077599,  by NSF Grants  IIS--0613435 and PHY--0650617, by the Swartz Foundation,
and by the Burroughs Wellcome Fund.
\end{acknowledgments}
\bibliographystyle{agu}
\bibliography{princeton}
\end{document}